\documentclass[useAMS,usenatbib]{mn2e}
\usepackage{graphicx}
\usepackage[latin1]{inputenc}
\usepackage{subfigure}
\usepackage{indentfirst}
\usepackage{multicol}
\usepackage{Times} 
\usepackage{url}
\usepackage{graphicx}
\usepackage{multicol}
\usepackage{epstopdf}

\title[On the location of the ring around the dwarf planet  Haumea]
{On the location of the ring around the dwarf planet Haumea}
\author[O. C. Winter, G. Borderes-Motta, T. Ribeiro]
{ O. C. Winter$^1$\thanks{E-mail: othon.winter@unesp.br}, 
G. Borderes-Motta$^1$\thanks{E-mail: gabriel\_borderes@yahoo.com.br},
T. Ribeiro$^1$\thanks{E-mail: taisalsiri@hotmail.com}\\ 
$^{1}$Grupo de Din\^amica Orbital e Planetologia, S\~ ao Paulo State University - UNESP, CEP 12516-410, Guaratinguet\'a, SP, Brazil
}

\begin{document}
%\date{Accepted 2008 January 21. Received 2008 January 21; in original form 2008 January 21}
\date{}
\pagerange{\pageref{firstpage}--\pageref{lastpage}} \pubyear{2017}
\maketitle
\label{firstpage}

%%%%%%%%%%%%%%%%%%%%%%%%%%%%%%%%%%%%%%%%%%%%%%%%%%%%%%%%%%%%%%%%%%%%%%%%%%%%%%%%%%%%%
\begin{abstract}
The recently discovered ring around the dwarf planet (136108) Haumea is located near the 1:3 resonance between the orbital motion of the ring particles and the spin of Haumea. In the current work is studied the dynamics of individual particles in the region where is located the ring. Using the Poincaré Surface of Section technique, the islands of stability associated with the 1:3 resonance are identified and studied. Along all its existence this resonance showed to be  doubled, producing pairs of periodic and quasi-periodic orbits. The fact of being doubled introduces a separatrix, which generates a chaotic layer that significantly reduces the size of the stable regions of the 1:3 resonance. The results also show that there is a minimum equivalent eccentricity ($e_{1:3}$) for the existence of such resonance. This value seems to be too high to keep a particle within the borders of the ring. On the other hand, the Poincaré Surface of Sections show the existence of  much larger stable regions, but associated with a family of first kind periodic orbits. They exist with equivalent eccentricity values lower than $e_{1:3}$, and covering a  large radial distance, which encompasses the region of the Haumea's ring. Therefore, this analysis suggests the Haumea's ring is in a stable region associated with a first kind periodic orbit instead of the 1:3 resonance.
\end{abstract}


%%%%%%%%%%%%%%%%%%%%%%%%%%%%%%%%%%%%%%%%%%%%%%%%%%%%%%%%%%%%%%%%%%%%%%%%%%%%%%%%%%
\section{Introduction}

(136108) Haumea is one of the four known dwarf planets beyond the orbit of Neptune.
It has two small satellites, an elongated shape and a short spin period ($T_{\rm Haumea}=3.9155$ hours).
Recently, through a multi-chord stellar occultation, Haumea was found to have a ring \citep{ortiz2017}. 
Data from this occultation combined with light curves constrained Haumea's triaxial ellipsoid shape, with values of semi-axes $a=1,161\pm 30$ km,
$b=852\pm 4$ km and $c=513\pm 16$ km \citep{ortiz2017}. 
The values of the ratios $a/b$ and $a/c$ are the highest found in bodies encircled by rings, 
rising the question on how such peculiar gravitational potencial would affect the ring particles.
The orbital radius of Haumea's ring  is  $r_{\rm ring}=2,287_{-45}^{+75}$ km, placing it close to the 1:3 resonance ($a_{1:3}=2,285\pm 8$ km) 
between the orbital motion of the ring particles and the spin of Haumea \citep{ortiz2017}.

In the present work we explore the dynamics of individual particles in the region where is located the ring.
The main goal is to identify the structure of the region in terms of location and size of the stable spots 
and also the reason for their existence.
Of particular interest is trying  to understand the dynamical structure associated with the 1:3 resonance. 
Considering a two-dimension system composed by the ellipsoidal Haumea and a massless particle in the equatorial plane, 
the Poincar\' e surface of section technique fits very well these objectives.

The adopted dynamical system and the Poincaré Surface of Section technique are presented in the next two sections. 
In Section 4 are discussed the characteristics of the resonant and non-resonant periodic orbits found in the surfaces of section. 
The locations and sizes of the stable regions associated with the 1:3 resonance and those associated with the first kind periodic orbits 
are compared to the location and size of  Haumea's ring in Section 5.
Our final comments are presented in the last section.

%%%%%%%%%%%%%%%%%%%%%%%%%%%%%%%%%%%%%%%%%%%%%%%%%%%%%%%%%%%%%%%%%%%%%%%%%%%%%%%%%%
\section{Dynamical System}

In the current work we study the trajectories of massless particles orbiting the ellipsoidal Haumea represented as an uniformly rotating  second degree and order gravity field.
The values of the gravitational potential coefficients that correspond to the oblateness ($C_{20}$) and to the ellipticity ($C_{22}$) are obtained from the ellipsoidal semi-axes using 
\citep{balmino1994}
\begin{equation}
C_{20}= (2c^2-a^2-b^2)/10{R_e}^2=-0.243
\label{eq:movx}
\end{equation}
and 
\begin{equation}
C_{22}=(a^2-b^2)/20{R_e}^2  =0.049
\label{eq:movx}
\end{equation}
where $R_e= (a\,b\,c)^{1/3}$ .

The equations of motion in the body-fixed frame (O$xy$), a frame uniformly rotating with the same period of Haumea's spin, as shown in Figure 1, is given by \citep{hu2004} 
\begin{equation}
\ddot{x} - 2 \omega \dot{y} = \omega ^{2} x + U_{x}$ ,$ 
\label{eq:movx}
\end{equation}
\begin{equation}
\label{eq:movy}
\ddot{y} + 2 \omega \dot{x} = \omega ^{2} y + U_{y}$ ,$
\end{equation}
where $\omega$ is the spin velocity of Haumea, $U_{x}$ and $U_{y}$ stand for the partial derivatives of the gravitational potential given in Equation (5). 
\begin{equation}
U(x,y)=\frac{G M}{r}\left( 1 - \left(\frac{R_{e}}{r} \right)^{2} \left[ \frac{C_{20}}{2}-\frac{3 C_{22}}{r^{2}}(x^{2}-y^{2})\right] \right),
\label{eq:}
\end{equation}
where $G$ is the gravitational constant, $M=4.006\times 10^{21}$ kg is  the mass of Haumea \citep{ortiz2017} and $r=\sqrt{x^2+y^2}$.
Conserved quantities, such as the Jacobi constant ($C_{j}$), can be useful to analyze the equations of motion. This constant is explicitly computed as \citep {hu2004}:
\begin{equation}
\label{eq-Cj}
C_{j} = \omega ^{2}(x^2+y^2)+2U(x,y)-\dot{x}^{2}-\dot{y}^{2}.
\end{equation}

Adopting the masses and orbits of
Namaka and Hi'iaka  \citep{ragozzine2009}, satellites of Haumea, we find that they have pericentre distances of about $20\times 10^{3}$ km and $47\times 10^{3}$ km,  
while their Hill radii are about $1\times 10^{3}$ km and $5\times 10^{3}$ km, respectively.
Therefore, the gravitational influence of these satellites on particles close to the location of the ring ($r\sim2\times 10^{3}$ km) is not significant. 
So, these two satellites are not taken into account in the current work.

\begin{figure}
\begin{center}
\includegraphics[scale=0.4]{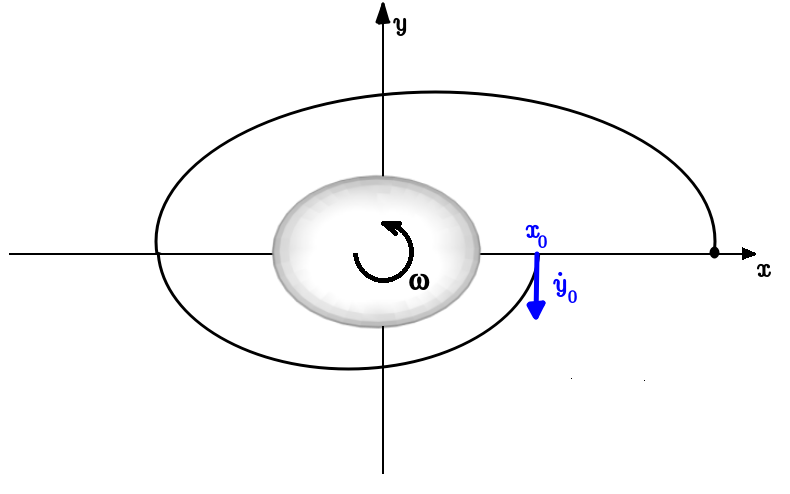}
\caption{Schematic diagram of a trajectory around Haumea fixed at a rotating $xy$ frame. $x_{0}$ marks the initial position of the trajectory and the blue arrow indicates the initial velocity.}
\label{fig_axe}
\end{center}
\end{figure}

%%%%%%%%%%%%%%%%%%%%%%%%%%%%%%%%%%%%%%%%%%%%%%%%%%%%%%%%%%%%%%%%%%%%%%%%%%%%%%%%%%
\section{Poincar\' e Surfaces of Section}

The Poincar\' e surface of section technique allows the determination of  the location and size of the regular and chaotic regions of low dimensional dynamical systems.
It also provides information on periodic and quasi-periodic orbits, resonant or not. This technique has been broadly applied in celestial mechanics.
With the development of computer machines, more than half century ago, they were used in many studies of the planar circular restricted three-body problem
 (H\' enon 1965a,b, 1966a, b, 1969; Jefferys 1971).

On the other hand, there are a few works that adopted the Poincar\'e surface of section when the dynamical system in study is a two-body problem, whose central body has non-spherical distribution of mass. For example, Scheeres et al. (1996) used it in order to find periodic orbits around the asteroid 4769 Castalia.
Jiang et al. (2016) showed the chaotic behaviour of trajectories around the asteroid 216 Kleopatra through  a couple of surfaces of section. 
\citet{borderes2018} also adopted it to explore the region nearby the asteroid 4179 Toutatis taking into account its three-dimensional irregular shape.

In the current work we follow the same approach described in \citet{borderes2018}.
The equations of motion (Equations (1-2)) are numerically integrated via the Bulirsh-Stoer integrator \citep{bulirsh1966}. 
The Poincar\'e surfaces of section are set in the plane $y=0$ and the initial conditions are systematically distributed over the $x$ axis. 
In general, it was fixed $y_{0}=\dot{x} _{0}= 0$, and $\dot{y}_{0}$ was computed for a fixed value of $C_{j}$ (Equation \ref{eq-Cj}), taking $\dot{y}_{0}<0$ (Figure 1).
During the integration, the conditions of the orbit are saved every time the trajectory crosses the section defined by $y=0$ with $\dot{y}<0$. 
The Newton-Raphson method is used to obtain an error of at most $10^{-13}$ from that section. 
The recorded points are plotted on the plane $(x,\dot{x})$ creating the Poincar\'e surface of section.

For a given value of $C_{j}$ is built one surface of section composed of many trajectories.
In general, we considered between 20 and 30 initial conditions for each $C_{j}$. 
In a surface of section, periodic orbits produce a number of fixed points, while quasi-periodic orbits generate islands (closed curves) around the fixed points.
On the other hand, chaotic trajectories are identified by points spread over a region, covering an area of the section. 

Generating surfaces of section for a wide range of $C_{j}$ values, we identified the islands associated with the 1:3 resonance.
These islands exist for $0.817\, {\rm km}^2{\rm s}^{-2}<C_{j}<0.828\, {\rm km}^2{\rm s}^{-2}$. 
The plots in Figure 2 show the whole evolution of the 1:3 resonance (in blue and green). 
Since the potential (Equation 5) is invariant under a rotation of $\pi$, formally, this 1:3 resonance is in fact a fourth order 2:6 resonance 
\citep{sicardy2018}, 
which is a doubled resonance that produces a pair of periodic orbits and their associated quasi-periodic orbits.
Therefore, in the Poincar\' e surfaces of section we have two pairs of mirrored sets of islands (one pair in blue and other in green), since
each periodic orbit generates two fixed points surrounded by islands of quasi-periodic orbits.

% Initially, we were expecting to see a pair of islands symmetric with respect to the line $\dot{x}=0$, 
% as was found in the 1:3 mean motion resonance  \citep{borderes2018}, since these are second order resonances.
%However, the family of periodic orbits associated with the 1:3 resonance is bifurcated, producing two pairs of asymmetric islands (in blue and green).

%Bifurcations of resonant periodic orbits in the restricted three-body problem are well known for a long time \citep{message1958, message1970} 
%and a short review on this topic can be found in \citet{winter1997b}.  
 
Due to the fact that this resonance is doubled, there is a separatrix between the two families of 
periodic/quasi-periodic orbits in the phase space. Consequently, there is a chaotic region generated by this separatrix, 
whose size changes according to the value of the Jacobi constant.  
% two islands into four smaller ones, introducing  separatrix and, consequently, more chaos in the region.  
%Therefore, this bifurcation reduces the size of the stability region produced by the resonance.
The quasi periodic orbits associated with the 1:3 resonance are indicated by the pairs of islands in blue and the pairs in green (Figure 2).    
For $C_{j}=0.818\, {\rm km}^2{\rm s}^{-2}$ these islands are relatively small and far from the centre of the red islands.
As the value of $C_{j}$ increases they increase in size and get closer to the centre of the red islands.
The blue and green islands reach the largest size at $C_{j}\sim 0.826\, {\rm km}^2{\rm s}^{-2}$, when they get very close to the red islands.
 With the increase of the $C_{j}$ value, the islands of the 1:3 resonance start shrinking as they get closer to the centre of the red islands, 
while they are surrounded by red islands ($C_{j}=0.827\, {\rm km}^2{\rm s}^{-2}$).
The evolution ends when the islands get too small and too close to the center of the red islands.
The red islands are quasi-periodic orbits associated with a family of periodic orbits of the first kind \citep{poincare1895}, which will be discussed in 
more detail in the next section.

\begin{figure*}
\begin{center}
\subfigure{\includegraphics[scale=0.38]{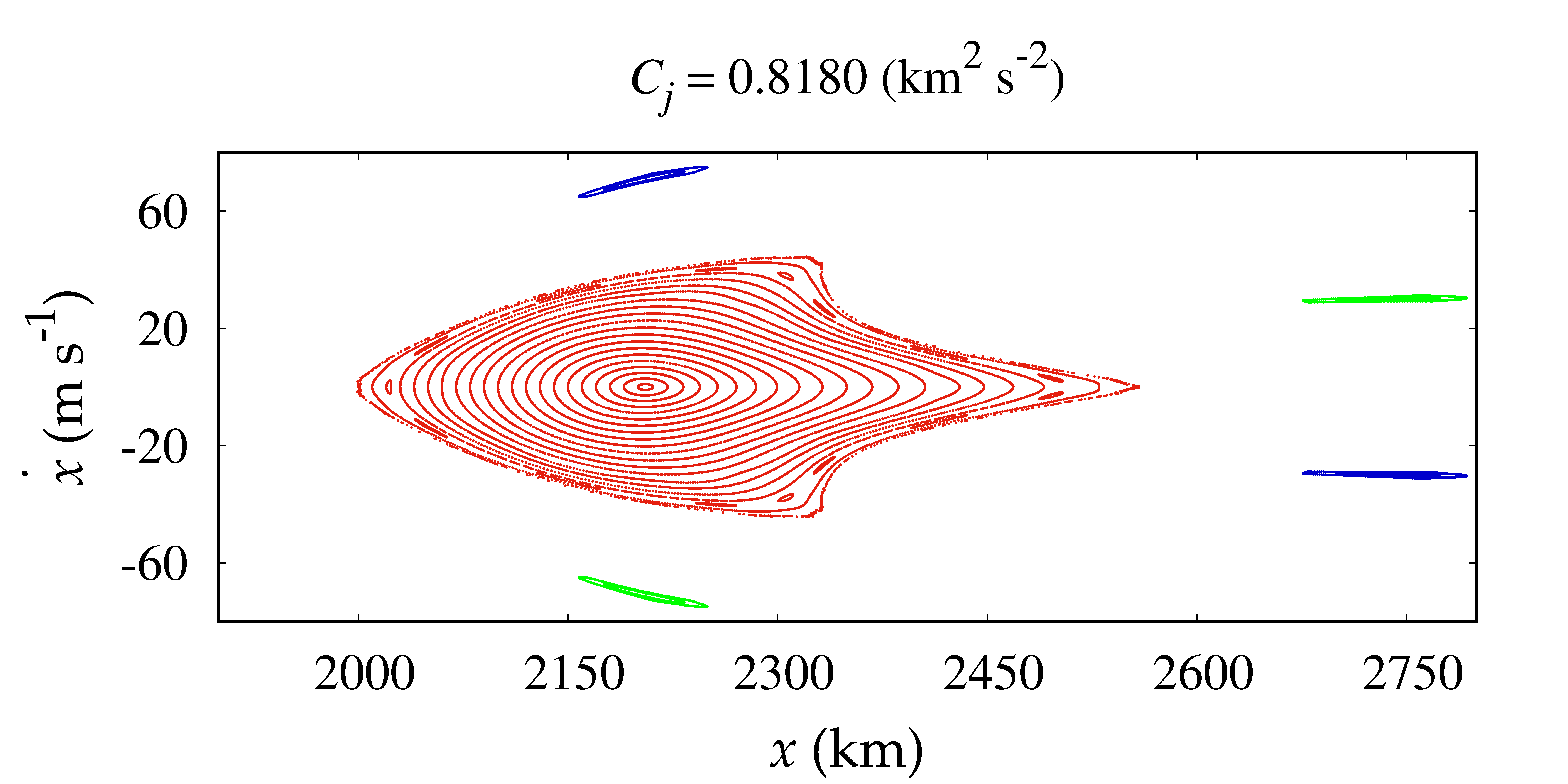}}
\subfigure{\includegraphics[scale=0.38]{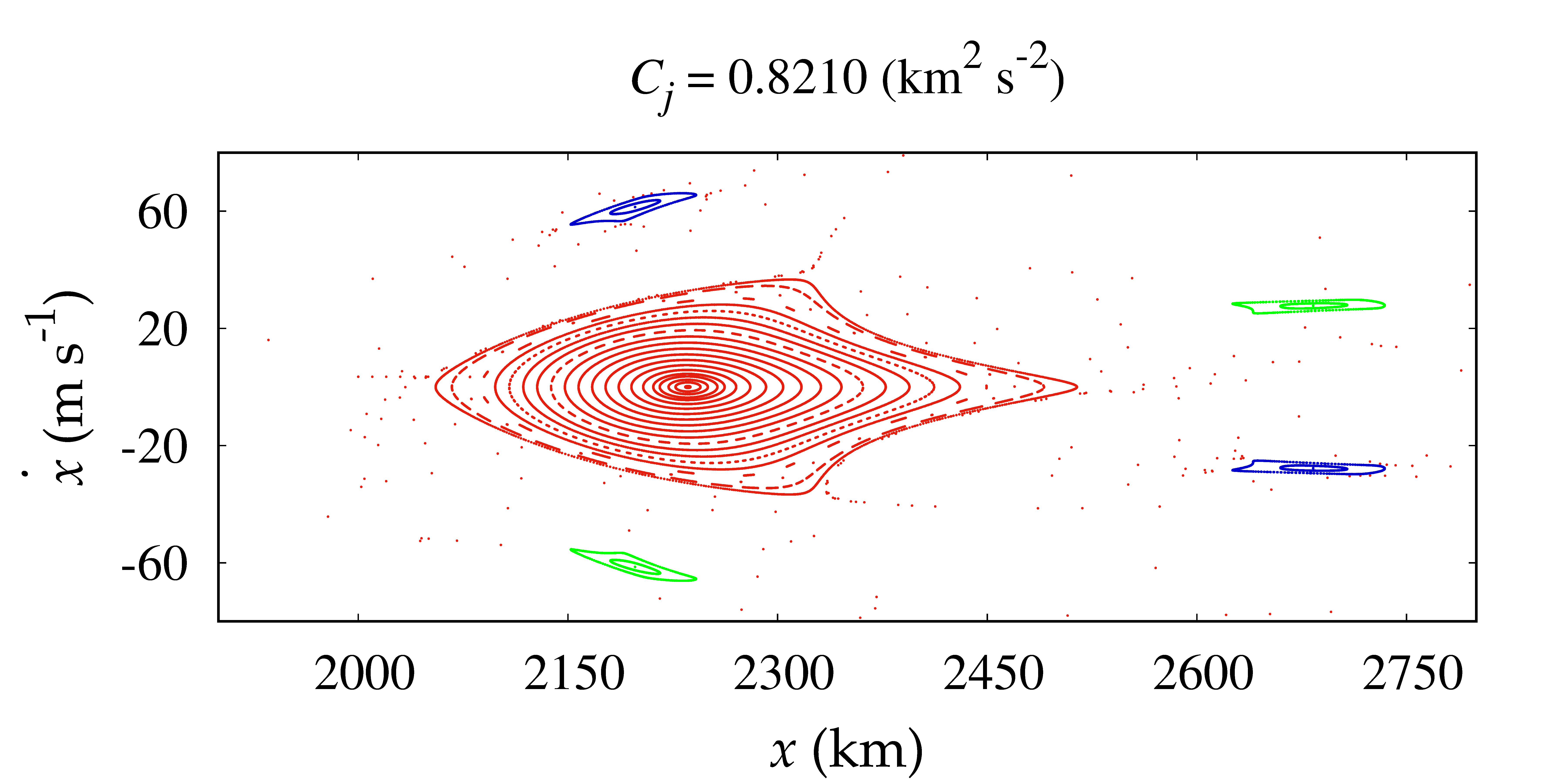}}
\subfigure{\includegraphics[scale=0.38]{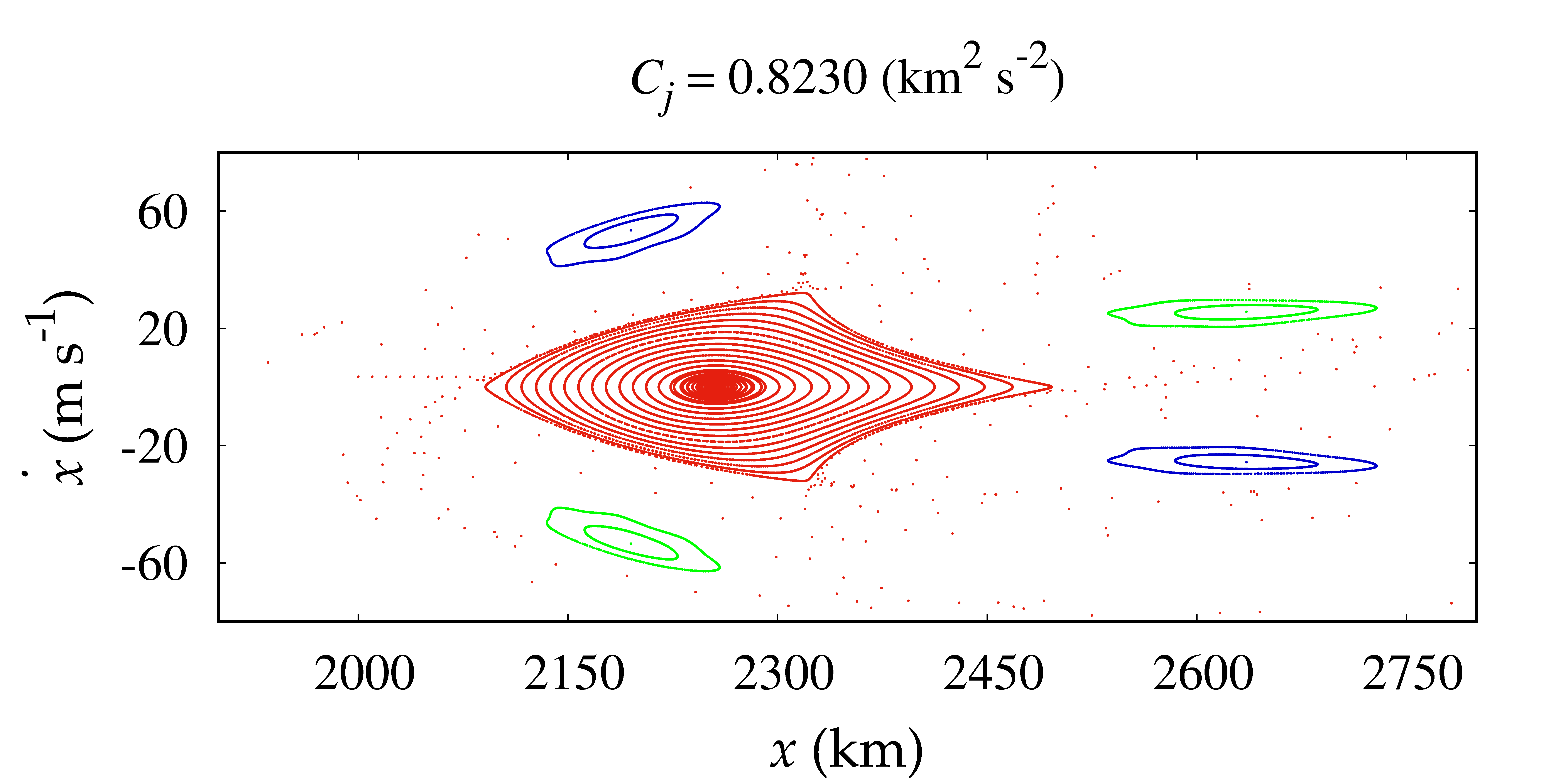}}
\subfigure{\includegraphics[scale=0.38]{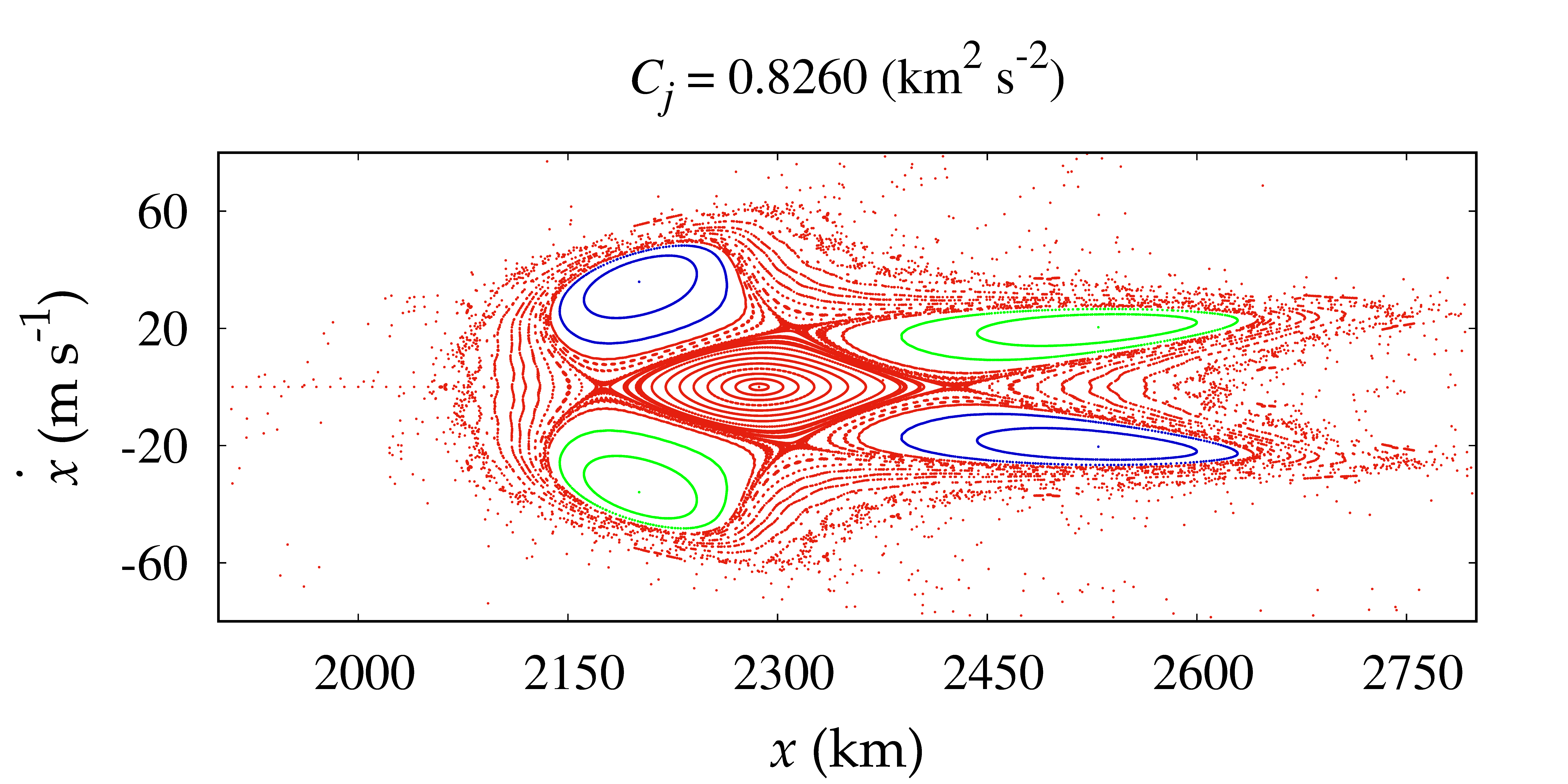}}
\subfigure{\includegraphics[scale=0.38]{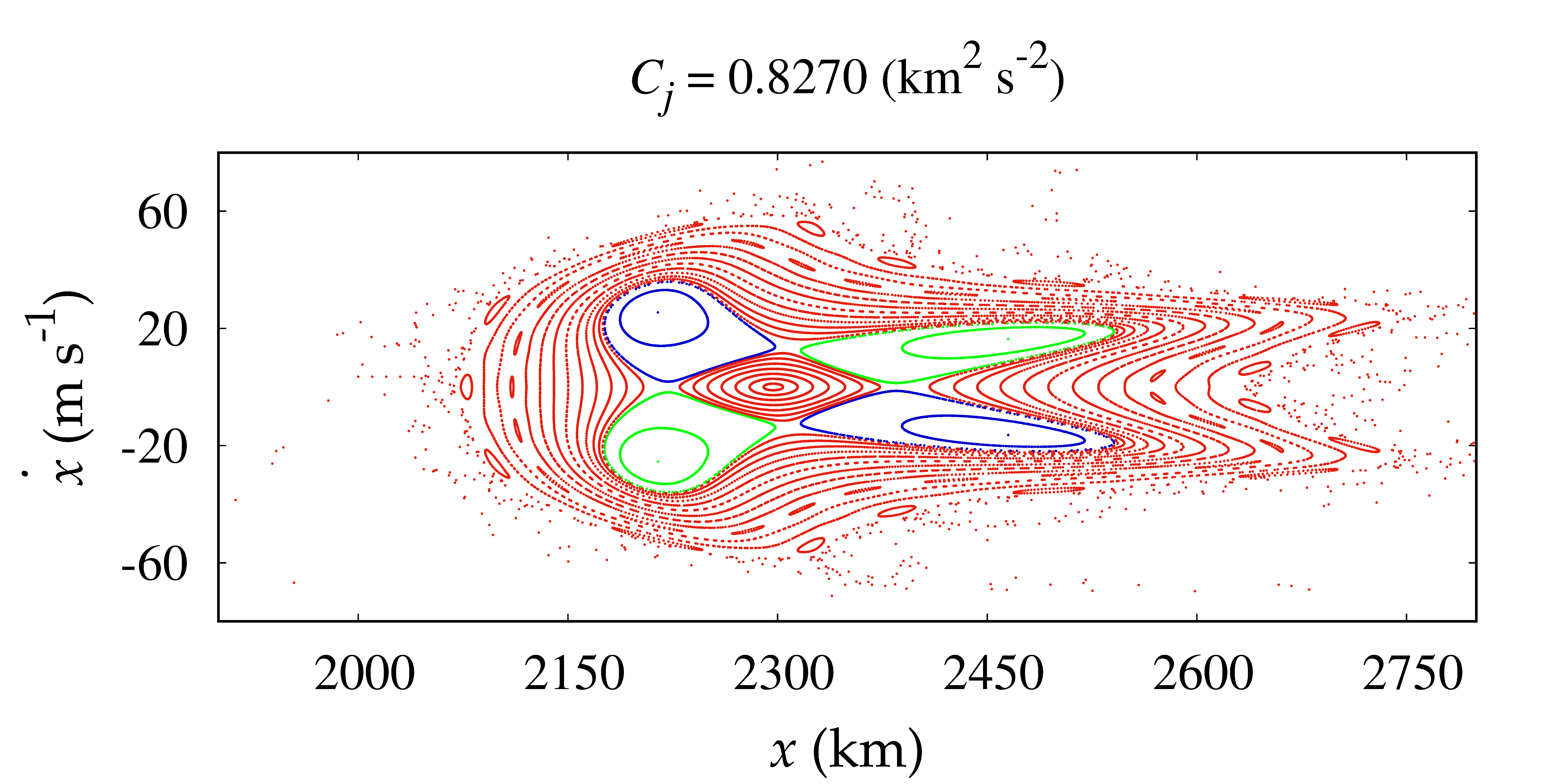}}
\subfigure{\includegraphics[scale=0.38]{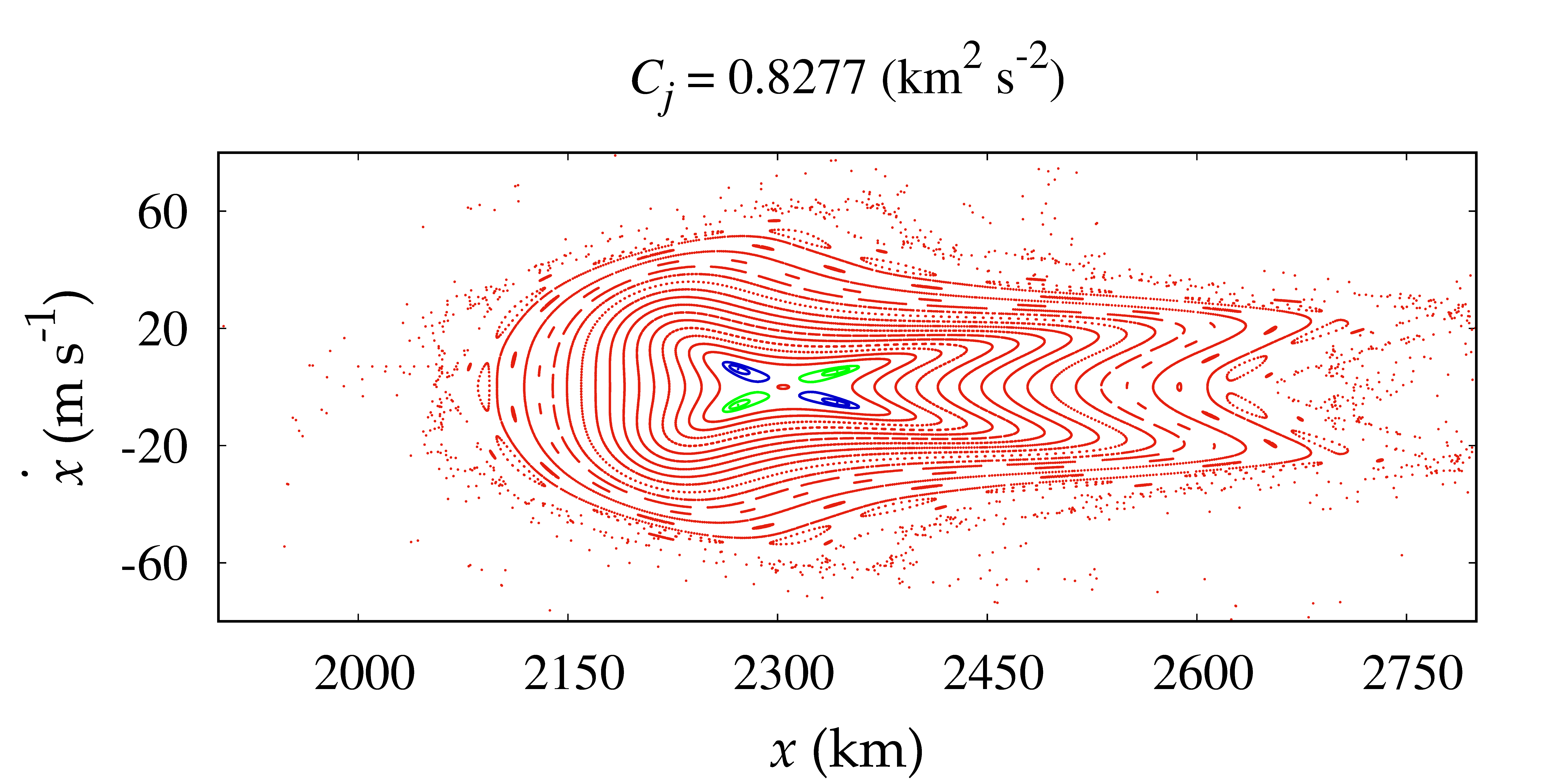}}
\caption{Poincar\' e surfaces of section for six different values of Jacobi constant ($C_j$) showing the whole evolution of the 1:3 resonance. 
The family of periodic orbits associated with this resonance are doubled into two families. One is indicated in blue and the other in green. 
The islands in red are quasi-periodic orbits  associated with a family of periodic orbits of the first kind. }
\label{fig_PSS}
\end{center}
\end{figure*}

%%%%%%%%%%%%%%%%%%%%%%%%%%%%%%%%%%%%%%%%%%%%%%%%%%%%%%%%%%%%%%%%%%%%%%%%%%%%%%%%%%
\section{Periodic Orbits}

In the planar, circular, restricted three-body problem, periodic orbits can be classified as periodic orbits of the {\it first kind} and of the {\it second 
kind} \citep{poincare1895, szebehely1967}.
The periodic orbits of the first kind are those originated from particles initially in circular orbits.
Considering the Earth-Moon mass ratio, \citet{broucke1968} explored the stability and evolution of several families of periodic orbits of the first kind.
A family of them was also studied for the Sun-Jupiter mass ratio in \citet{winter1997a}. 
More recently, motivated by the New Horizons mission, \citet{giuliatti2014} identified a peculiar stability region between Pluto and Charon, called {\it 
Sailboat island}, 
and showed that it was associated with a family of periodic orbits of the first kind.
In the case of periodic orbits of the second kind, the particles are in eccentric orbits in a mean motion resonance, the so-called {\it resonant periodic 
orbits}.

Now, considering the restricted two-body problem, where the primary is a rotating non-spherical triaxial body, \citet{borderes2018} showed  
examples of both kinds of periodic orbits.
Among them, they identified resonant periodic orbits as the 3:1 resonance between the mean motion of the particle and the spin of the central body, and 
one periodic orbit of the first kind.

As seen in the previous section, the Poincar\' e surfaces of section showed that the dynamical structure of the  region where is located the 
Haumea's ring is determined by the 1:3 resonant periodic orbits and also by a family of periodic orbits of the first kind.
In order to better understand these periodic orbits we will explore some of their features.

Since the 1:3 resonant periodic orbits are doubled, we selected a pair of them to study.  
Considering the Jacobi constant $C_j=0.826\, {\rm km}^2{\rm s}^{-2}$, Figure 3 shows the pair of 1:3 resonant periodic orbits. 
These are the periodic orbits shown in the Poincar\' e surface of section of Figure 2 (second row and second column), as the points located at the 
centre of the islands in blue, and those at the centre of the islands in green.   

\begin{figure*}
\begin{center}
%\subfigure{\includegraphics[scale=0.68]{res-T297-0818-sg.eps}}
%\subfigure{\includegraphics[scale=0.68]{res-T297-0818-sg-b.eps}}
%\subfigure{\includegraphics[scale=0.68]{res-T297-0818-si.eps}}
%\subfigure{\includegraphics[scale=0.68]{res-T297-0818-si-b.eps}}
\subfigure{\includegraphics[scale=0.68]{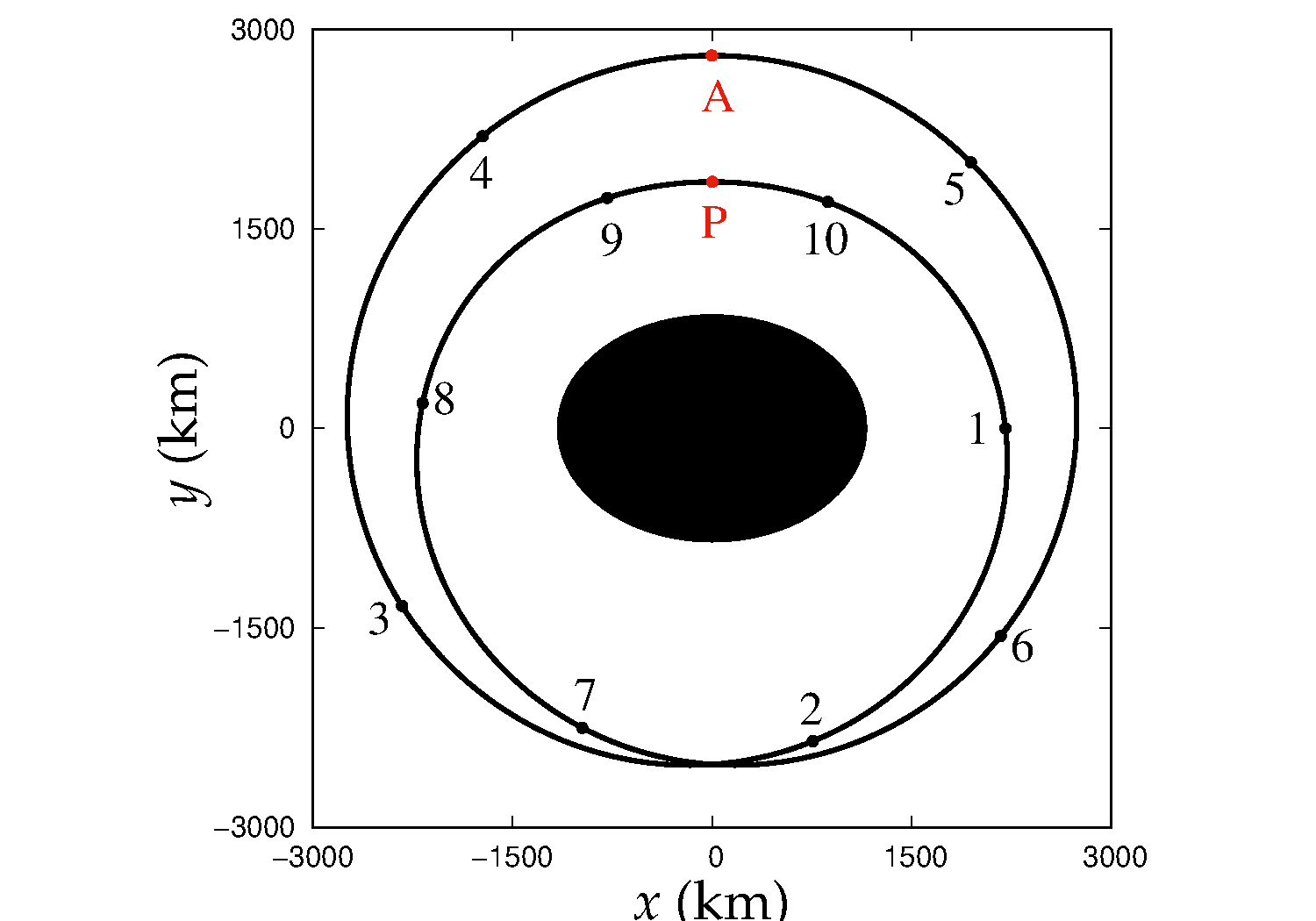}}
\subfigure{\includegraphics[scale=0.68]{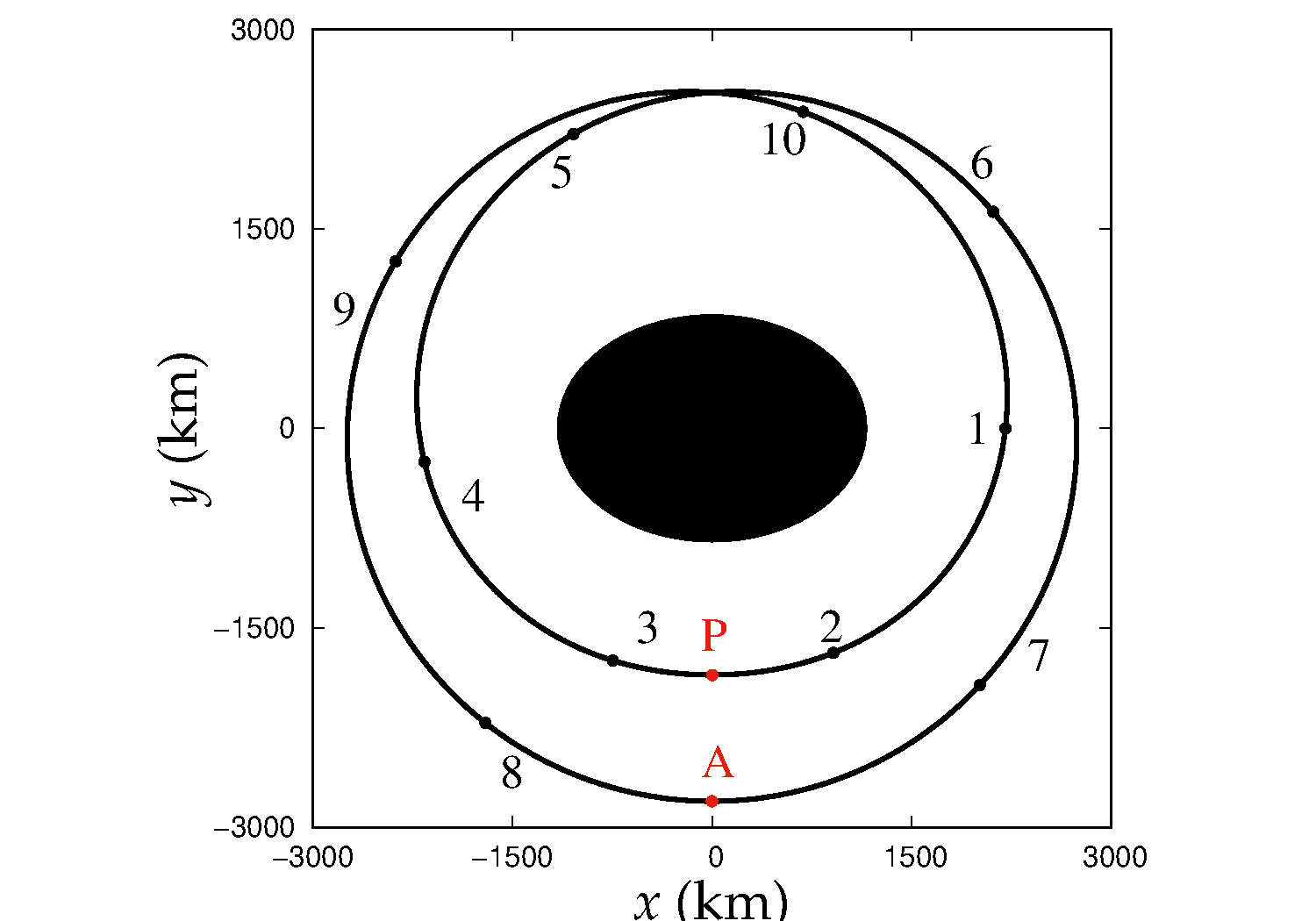}}
\subfigure{\includegraphics[scale=0.68]{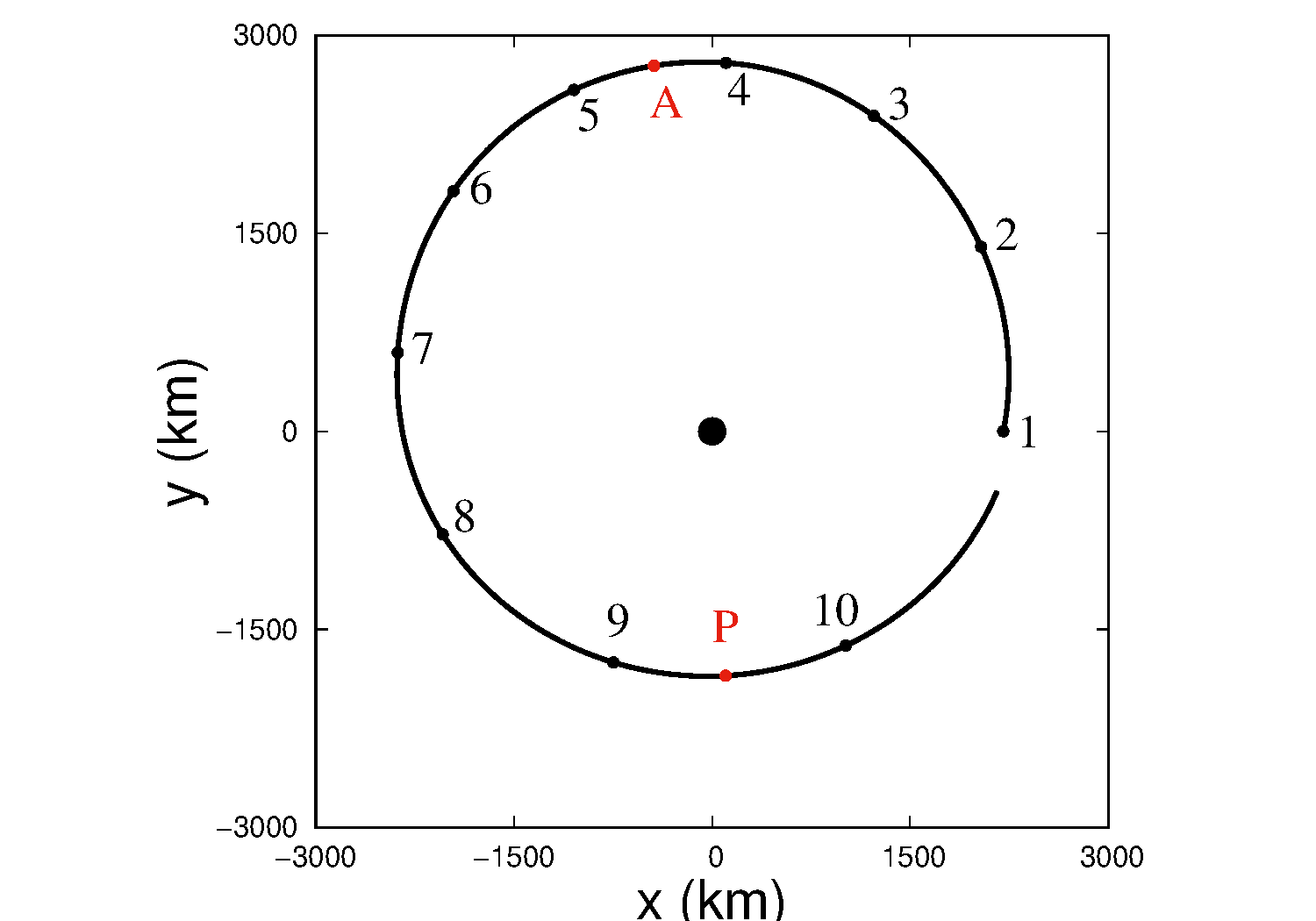}}
\subfigure{\includegraphics[scale=0.68]{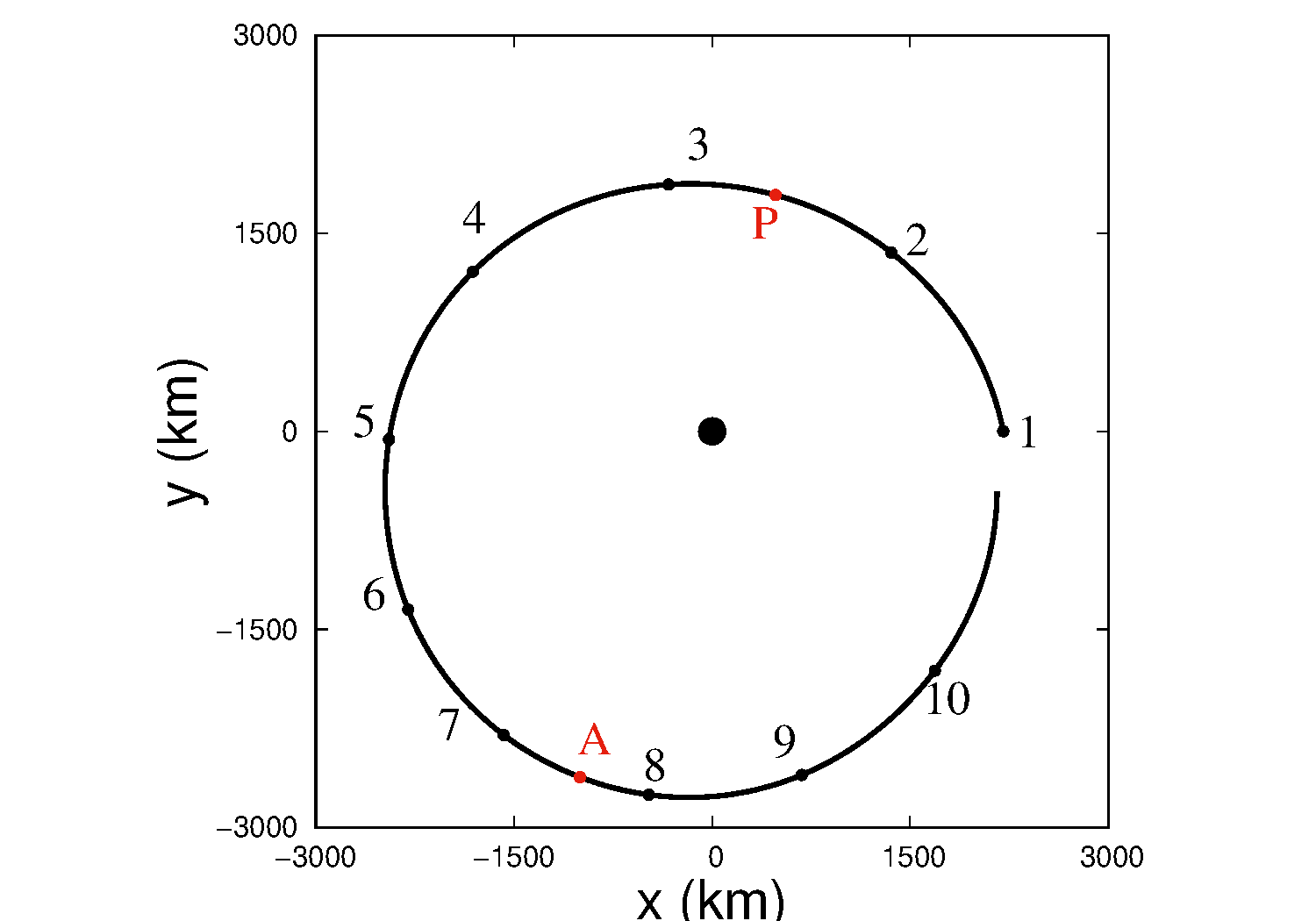}}
\subfigure{\includegraphics[scale=0.35]{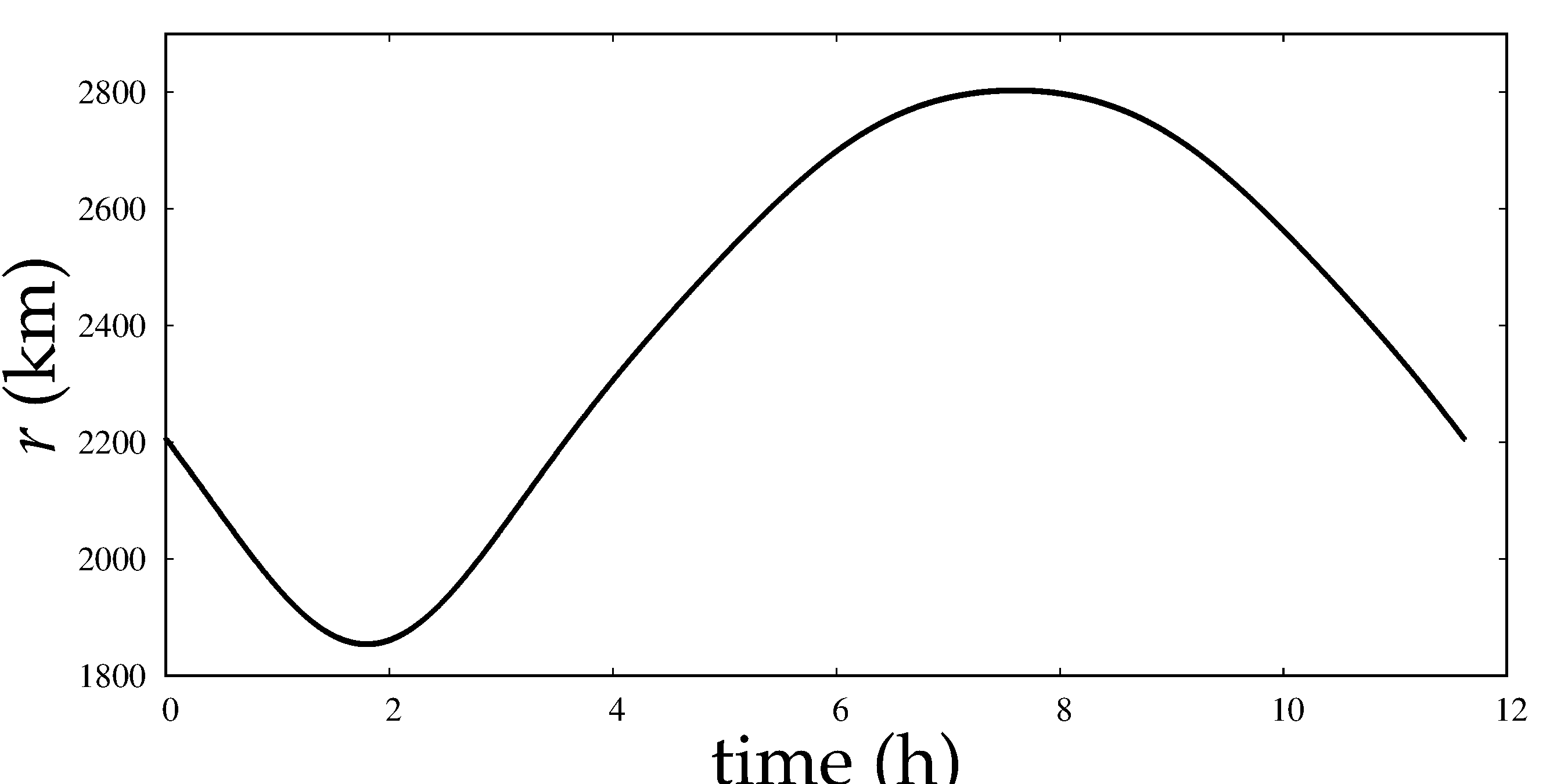}}\qquad\qquad\qquad
\subfigure{\includegraphics[scale=0.35]{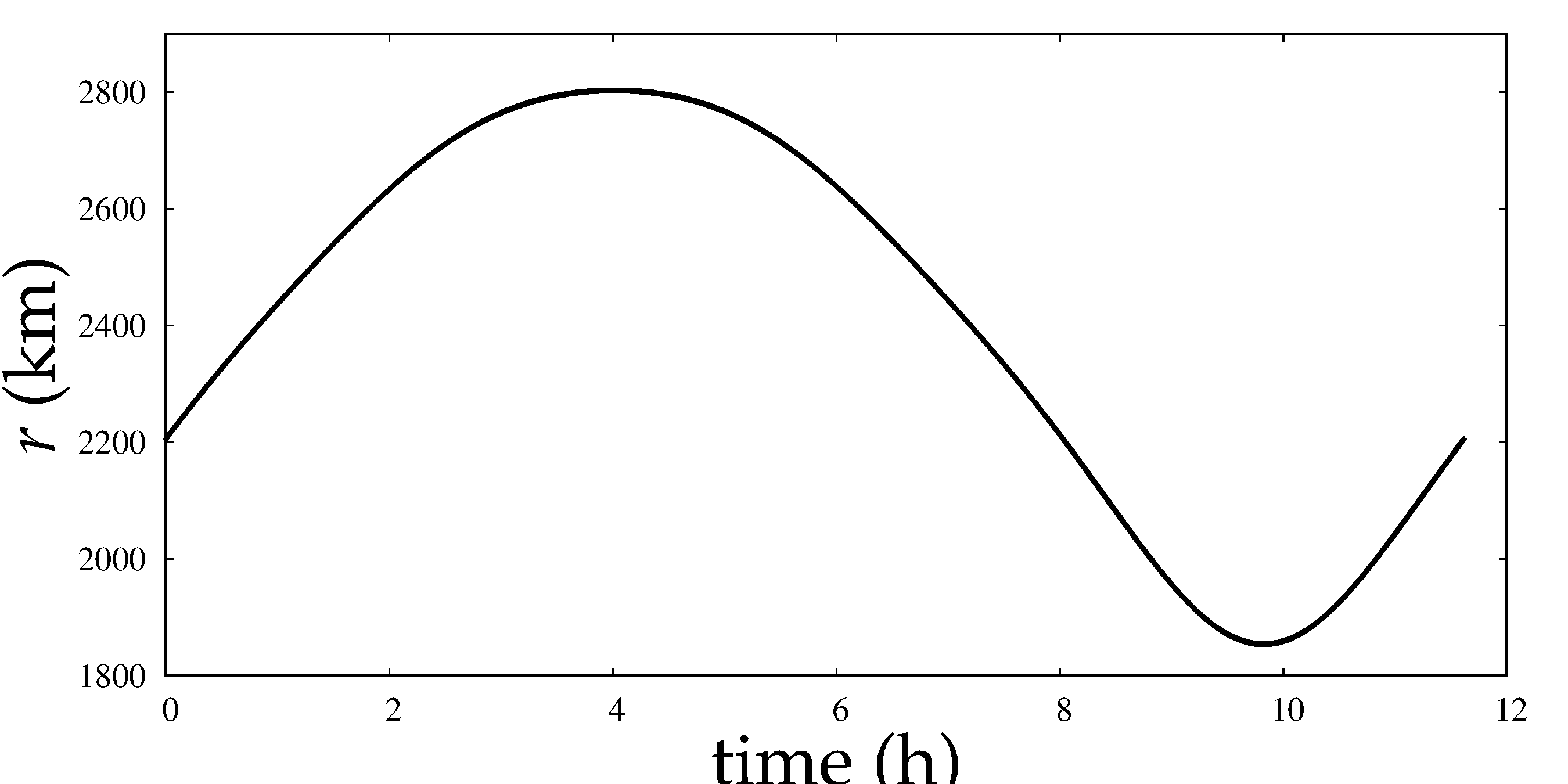}}
\caption{Example of a pair of 1:3 resonant periodic orbits whose Jacobi constant value is $C_j=0.826\, {\rm km}^2{\rm s}^{-2}$.  
They are the periodic orbits shown in the Poincar\' e surface of section of Figure 2 (second row and second column).
The points at the centre of the islands in green correspond to the periodic orbit shown  in the plots of the first column,
while those at the centre to the islands in blue correspond to the periodic orbit shown  in the plots of the second column.   
In the first row are plotted the trajectories in the rotating frame, while  trajectories in the inertial frame are plotted in the second row. 
The temporal evolution of the orbital radii are shown in the last row. The numbers in the plots of the first and second rows show locations equally 
spaced in time, indicating the time evolution of the trajectories. The letters P and A indicate the location of the pericentre and the apocentre.}
\label{fig_orb}
\end{center}
\end{figure*}

The trajectories in the rotating frame are plotted in the first row. 
The numbers indicate the time evolution of the trajectories and show the locations equally spaced in time.
In the rotating frame both trajectories are retrograde and symmetric with respect to the line $x=0$, where are located the pericentre (P) and 
the apocentre (A) of each trajectory. 
One trajectory is a mirrored image of the other with respect to the line $y=0$. 
That is a natural consequence of the potential (Equation 5) being invariant under a rotation of $\pi$.
The period of these periodic orbits corresponds to three periods of  Haumea's rotation.
In the inertial frame, the trajectories are prograde (second row).    
Note that these trajectories are also periodic in the inertial frame. 
The temporal evolution of the orbital radii  of both trajectories (last row) also helps to better visualize the trajectories shape. 

The 1:3 resonant angle is given by
$$
\phi_{1:3}= \lambda_H - 3\lambda + 2\varpi \, ,
$$   
where  $\lambda$ and $\varpi$  are the mean longitude and longitude of periapse of the particle, respectively, and $\lambda_H$  is Haumea's 
orientation angle.
For all resonant trajectories shown in Figure 2, we verified that the resonant angle oscillates around $\pi$ (orbits in blue) or $-\pi$ (orbits in green).

%Figure 4 shows the values of $\phi_o$ for the 1:3 resonant periodic orbits given in Figure 2 as a function of their Jacobi constant value. 
%Note that the value of $\phi_o$ reduces as the Jacobi constant increases.

%\begin{figure}
%\begin{center}
%\includegraphics[scale=0.37]{po-eq-li.eps}
%\caption{ The constant values of $\phi_{1:3}=\phi_o$ for the resonant periodic orbits shown in Figure 2 as a function of their Jacobi constant.}
%\label{fig_axe}
%\end{center}
%\end{figure}

An example of a first kind periodic orbit, with $C_j=0.826\, {\rm km}^2{\rm s}^{-2}$, is presented in Figure 4.
The shape of the trajectory in the rotating frame follows similarly  the shape of Haumea. 
The furthest points of the trajectory are aligned with the long axis, while the closest points are alined with the short axis.
The period of this periodic orbit is $T= 5.979$ h, slightly more than 1.5 times the spin period of Haumea.
Note that, at the same time this trajectory completes one period in the rotating frame, it completed only about half of its orbit around Haumea in the 
inertial frame. 
The temporal evolution of the radial distance may mislead to the conclusion that the trajectory has a pair of pericentre and a pair of apocentre.
However, this trajectory is not a usual keplerian ellipse, with the central body at one of the foci.
In fact, the trajectory looks like an ellipse with the central body at its centre.   

A comparison between the radial amplitudes of oscillation of this periodic orbit of the first kind and the resonant periodic orbits given in Figure 3  
shows a huge difference. While the periodic orbit of the first kind oscillates less than 30 km, the resonant periodic orbit shows a radial oscillation of 
almost 600 km.
This is a striking information that will be analysed for the whole set of periodic orbits in the next section. 

\begin{figure}
\begin{center}
\subfigure{\includegraphics[scale=0.68]{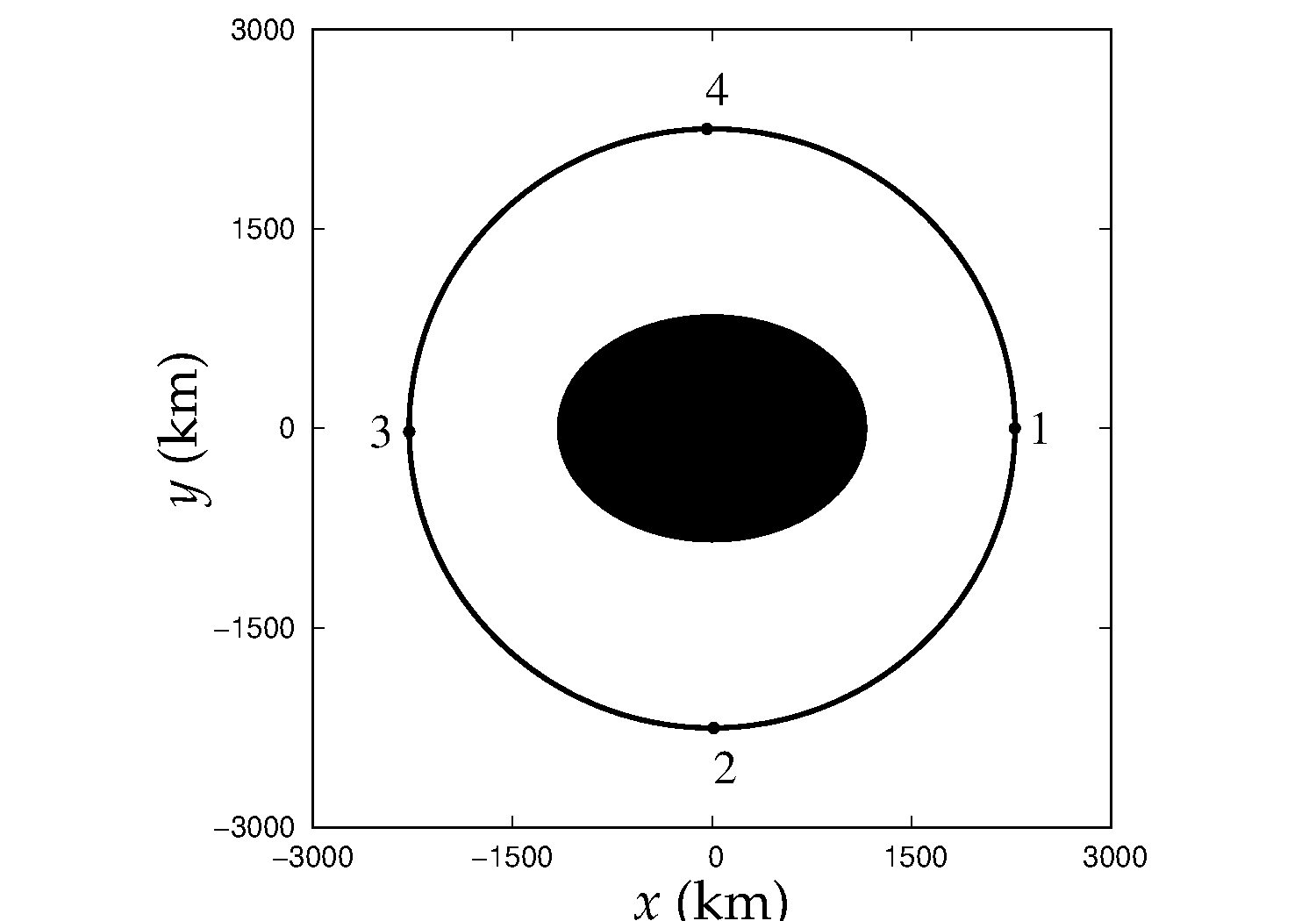}}
\subfigure{\includegraphics[scale=0.68]{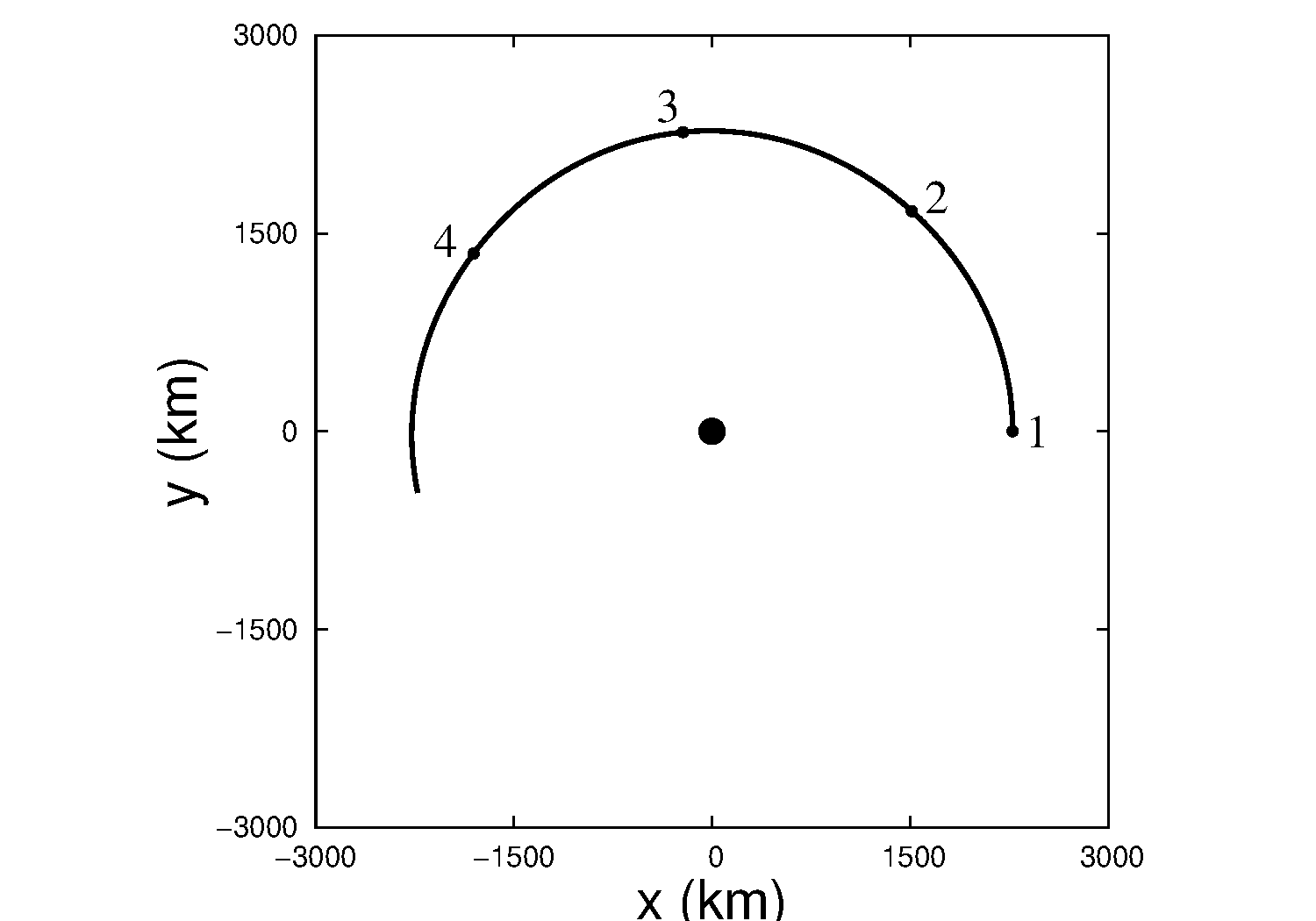}}
\subfigure{\includegraphics[scale=0.35]{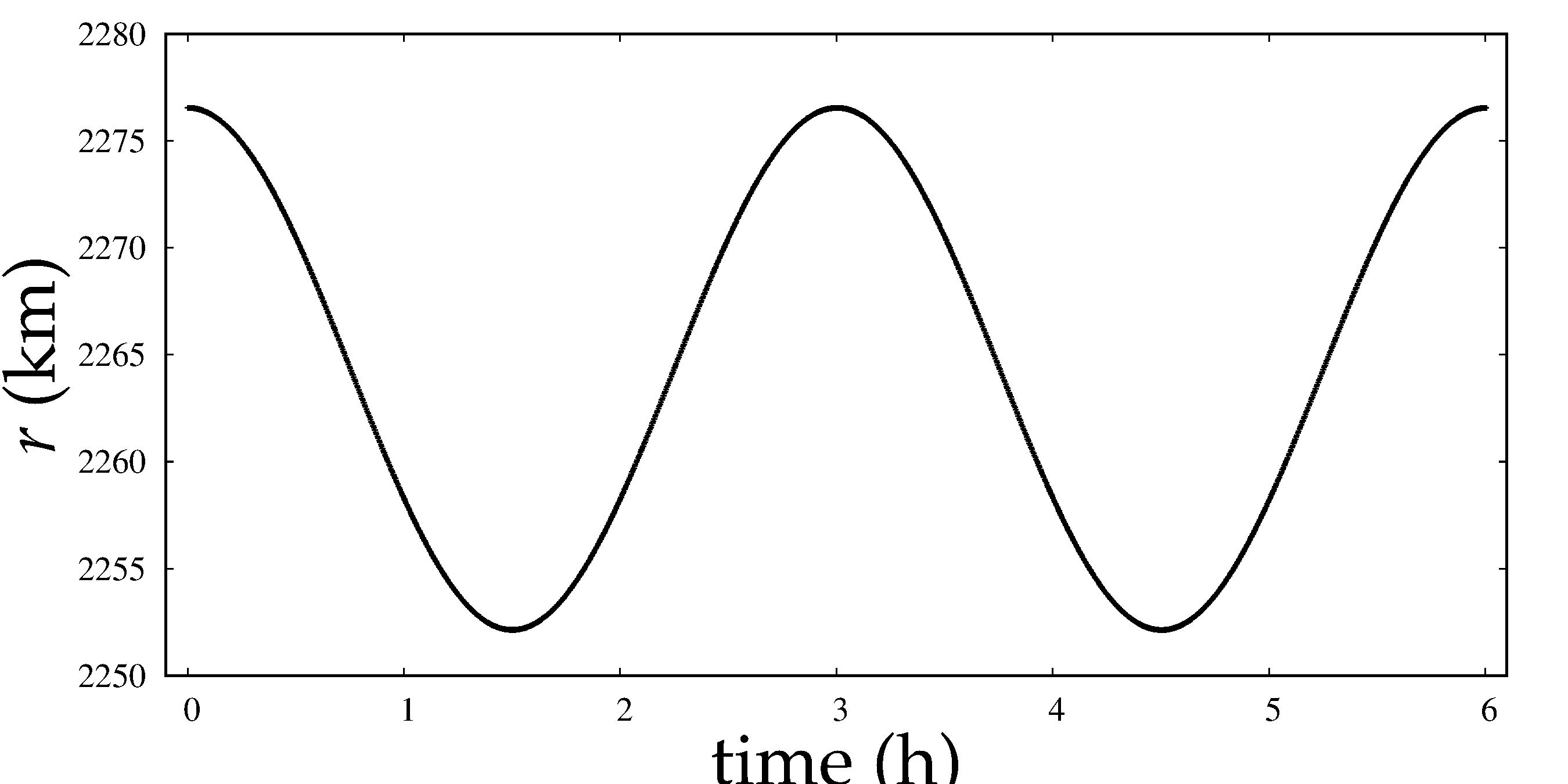}}
\caption{Example of a first kind periodic orbit with $C_j=0.826\, {\rm km}^2{\rm s}^{-2}$ and period of the periodic orbit $T= 5.979$ h. 
In the top plot is shown the trajectory in the rotating frame, while the trajectory in the inertial frame is in the middle plot. 
The temporal evolution of the orbital radius is shown in the bottom plot.
The numbers inside the top and middle plots show locations equally spaced in time,
 indicating the time evolution of the trajectory.}
\label{fig_orb}
\end{center}
\end{figure}

One feature that makes the difference between a resonant periodic orbit and a periodic orbit of the first kind is that the period (in the rotating frame) 
of a periodic orbit of the first kind varies significantly, in a range that might cross several values that are commensurable without showing a resonant 
behaviour. 
Figure 5 shows the period of the periodic orbits  in the rotating frame as a function of their Jacobi constant, $C_j$. 
While the resonant periodic orbits 1:3 (green) exists only nearby the period commensurable with the spin period of Haumea,
the period of the periodic orbits of the first kind cover a wide range of values (red).
It passes through several periods that, in the rotating frame, are commensurable with the spin period of Haumea.

\begin{figure}
\begin{center}
\subfigure{\includegraphics[scale=0.65]{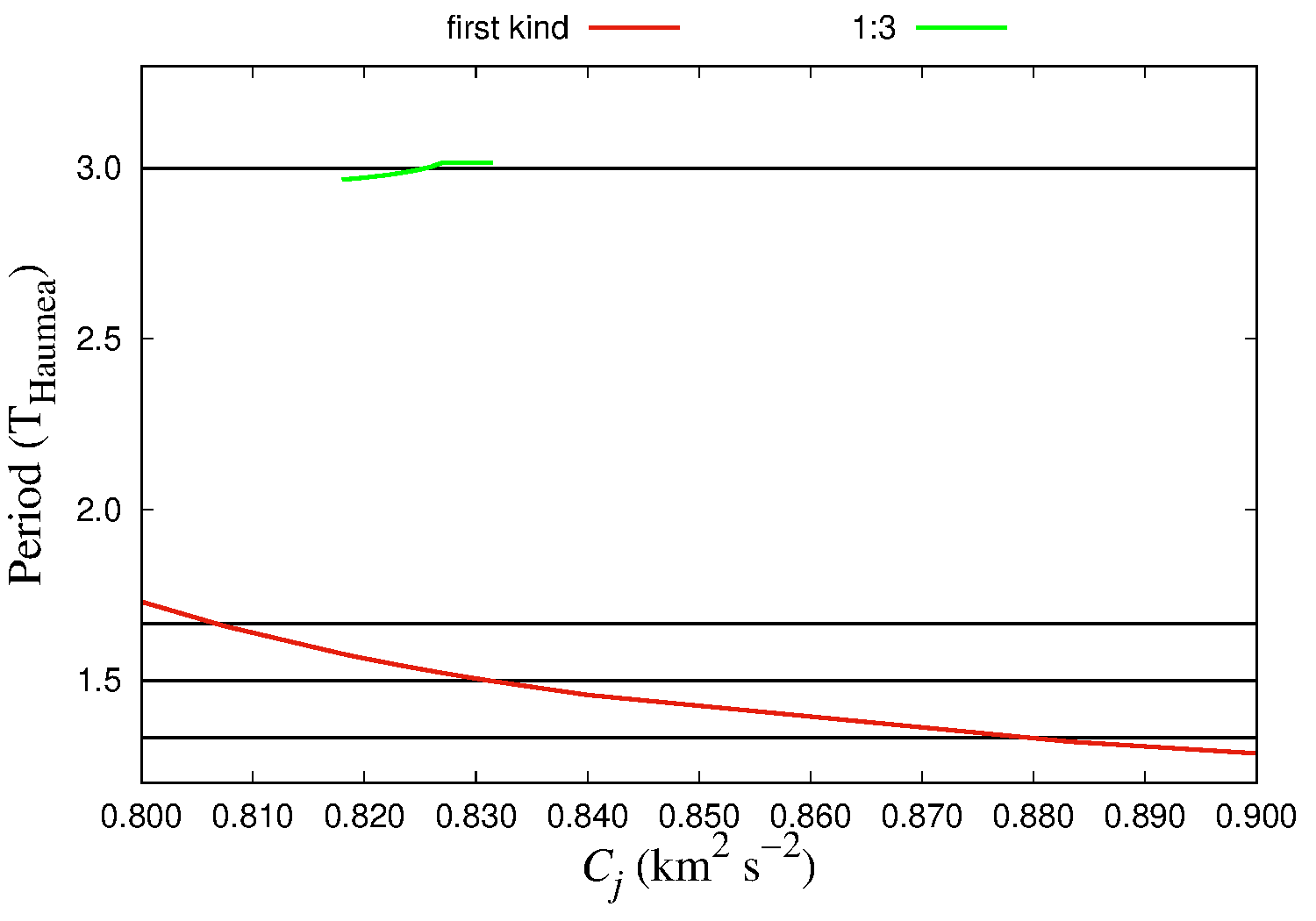}}
\subfigure{\includegraphics[scale=0.65]{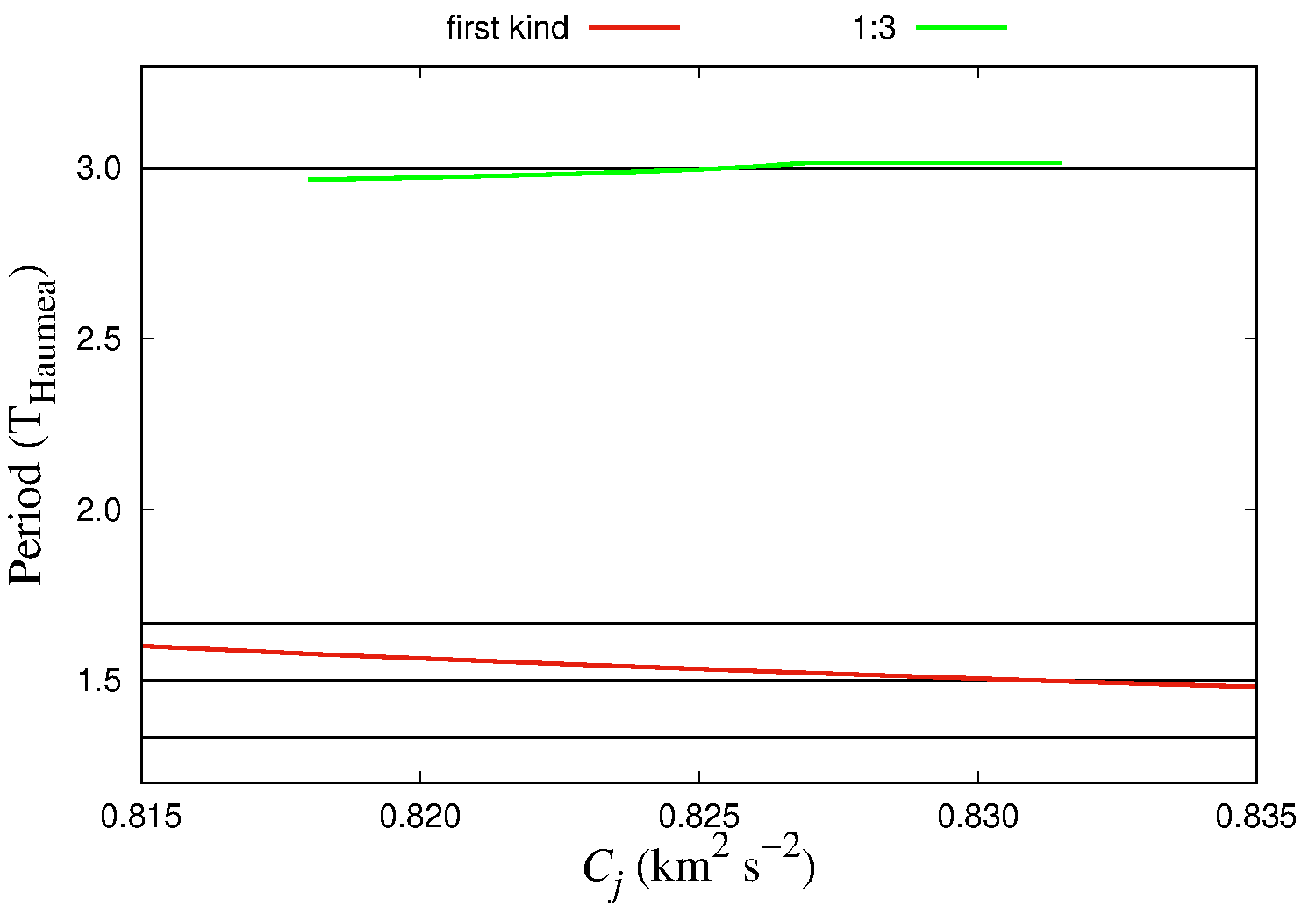}}
\caption{Period of the periodic orbits in the rotating frame as a function of their Jacobi constant, $C_j$. In green are the values for the 1:3 resonance 
and in red those for the periodic orbits of the first kind. 
The black lines indicate some periods that, in the rotating frame, are commensurable with the spin period of Haumea. The bottom plot is a zoom of the top plot.}
\label{fig_rtd}
\end{center}
\end{figure}

%%%%%%%%%%%%%%%%%%%%%%%%%%%%%%%%%%%%%%%%%%%%%%%%%%%%%%%%%%%%%%%%%%%%%%%%%%%%%%%%%%
\section{The Location of the Ring}

In this section we work on a correlation between the locations of the stable regions associated with the periodic orbits and the location of Haumea's 
ring.

The periodic orbits of the first kind are like ellipses with Haumea at its centre (top of Figure 4), not at one of the foci as in a usual Keplerian ellipse. 
Therefore, we will define an equivalent semi-major axis ($a_{\rm eq}$) and an equivalent eccentricity ($e_{\rm eq}$) as the values of a keplerian 
ellipse whose pericentre and apocentre are the same as those of the maximum ($r_{\rm max}$) and minimum ($r_{\rm min}$) radial distances of the 
trajectory to Haumea, respectively (as illustrated in Figure 6). Then, an ellipse with Haumea at one of the foci and with semi-major axis $a_{\rm eq}= 
(r_{\rm min}+r_{\rm max})/2$ and eccentricity $e_{\rm eq}= 1-(r_{\rm min}/a_{\rm eq})$ will cover the same radial distance range from Haumea as 
that  one covered by the first kind periodic orbit. Similarly, the Haumea's ring, whose nominal location \citep{ortiz2017} is given by an internal radius 
at  2252 km and an external radius at 2322 km, can be represented by a range of keplerian ellipses that fits such radial extent.

Figure 7 presents a diagram  $a_{\rm eq}$ vs. $e_{\rm eq}$ showing the locations of the Haumea's ring, the 1:3 resonant periodic orbits and the first 
kind periodic orbits.
Note that the largest possible eccentricity for a ring particle  ($e_{\rm eq}=0.0153$) is smaller than the smallest eccentricity for the 1:3 resonant 
periodic orbits  ($e_{\rm eq}=0.0166$). 
This means that all  periodic orbits of the 1:3 resonance show a radial variation that goes beyond the location of the Haumea's ring.  
On the other hand, first kind periodic orbits cover an interval of semi-major axes and eccentricities that fits inside the ring range values.
Therefore, these are indications that the first kind periodic orbits are more strongly connected to the Haumea's ring than the 1:3 resonance.

The results shown in Figure 7 consider only the periodic orbits (fixed points inside the islands in the Poincar\' e surfaces of section). However, the 
size of the stability regions around the periodic orbits is determined by the quasi-periodic orbits around them. As introduced in \citet{winter2000}, 
the size and location of the stability region associated with the first kind periodic orbits can be  determined from the Poincar\' e surfaces of section, 
for each Jacobi constant ($C_j$), by identifying the pair of $x$ values of the largest islands when $\dot x = 0$.  
This approach provides the maximum libration amplitude of the quasi-periodic orbits around the family of periodic orbits. 
Figure 8 shows such stability region in a diagram of  $C_j$ vs. $x$.  The stability region (in gray) encompasses the nominal radial region of the 
Haumea's ring (in blue)
being at least 200 km larger.  

Therefore, the stable region associated with the first kind periodic orbit family is at the right location, is wide enough, and presents trajectories with 
eccentricities low enough to fit within the borders of  Haumea's ring.

\begin{figure}
\begin{center}
\includegraphics[scale=0.14]{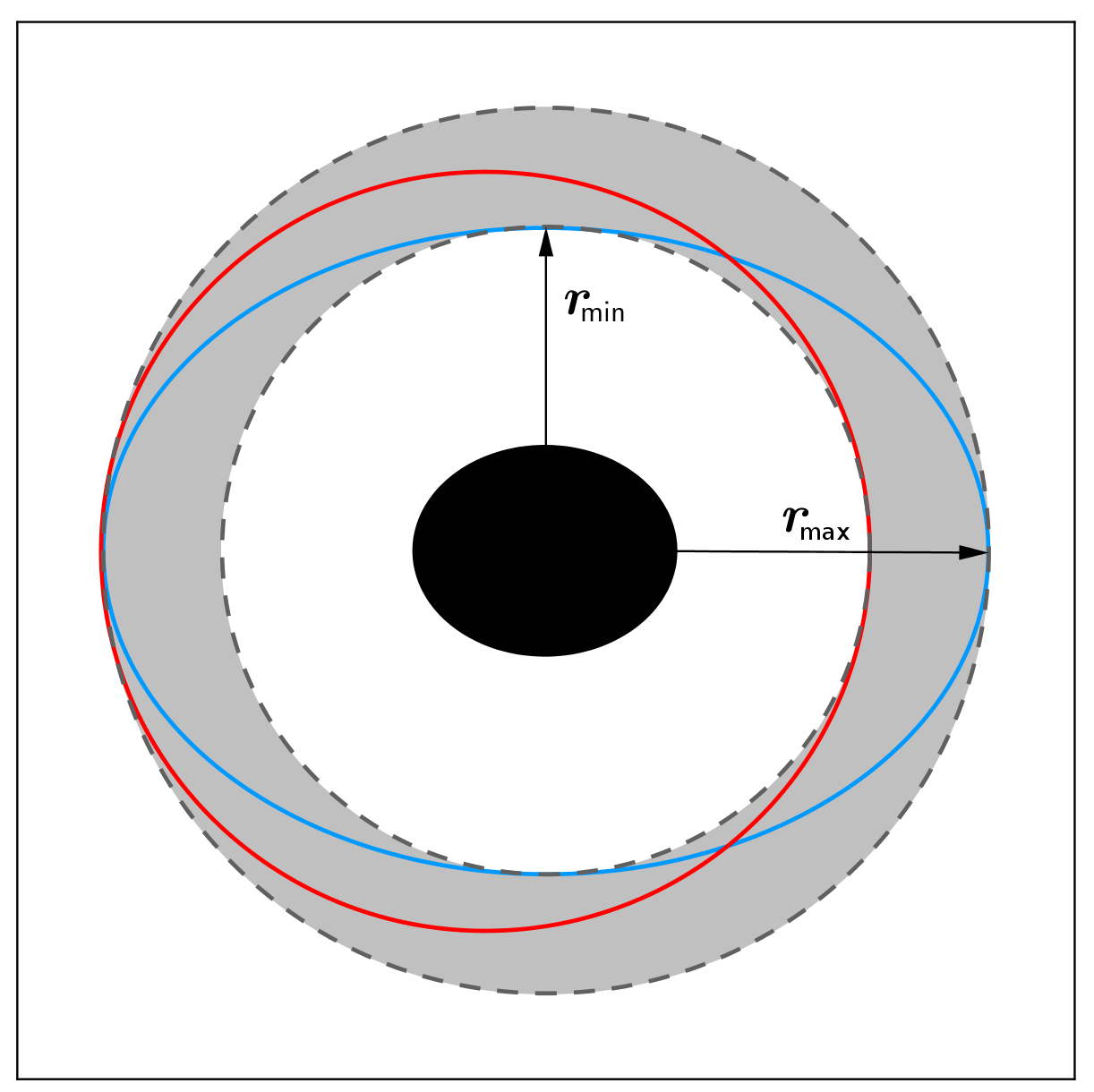}
\caption{Schematic diagram for the definition of $a_{\rm eq}$ and $e_{\rm eq}$. The periodic orbit in the rotating frame is an ellipse (in blue) 
centered at Haumea with semi-major axis $r_{\rm max}$ and semi-minor axis $r_{\rm min}$. This orbit covers a radial distance from Haumea 
varying between $r_{\rm min}$ and $r_{\rm max}$, indicated by the gray region. An ellipse (in red) with Haumea in one of the foci and with $a_{\rm 
eq}= (r_{\rm min}+r_{\rm max})/2$ and  $e_{\rm eq}= 1-(r_{\rm min}/a_{\rm eq})$ as semi-major axis and eccentricity also covers the same radial 
distance range from Haumea as the periodic orbit does (gray region).}
\label{fig_axe}
\end{center}
\end{figure}

\begin{figure}
\begin{center}
\includegraphics[scale=0.65]{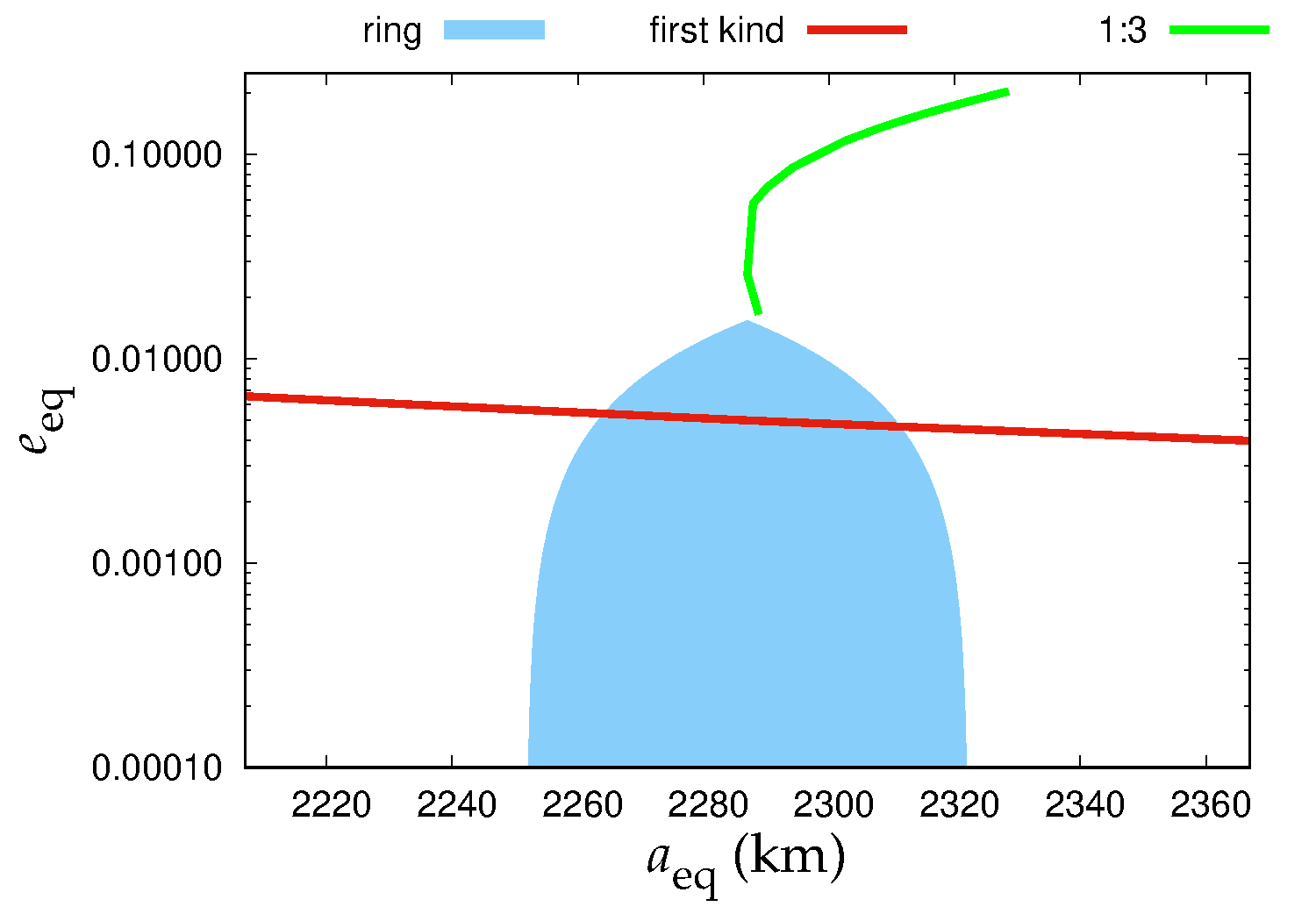}
\caption{Diagram of equivalent semi-major axis versus equivalent eccentricity, $a_{\rm eq}$ vs. $e_{\rm eq}$.  The green line indicates the values of  
($a, e$) for each periodic orbit associated with the 1:3 resonance. The red line indicates the values of  ($a_{\rm eq}, e_{\rm eq}$)
for each periodic orbit of the first kind. The blue region shows  the range of ($a_{\rm eq}, e_{\rm eq}$) values that corresponds to the location of 
Haumea's ring.}
\label{fig_axe}
\end{center}
\end{figure}

\begin{figure}
\begin{center}
\includegraphics[scale=0.65]{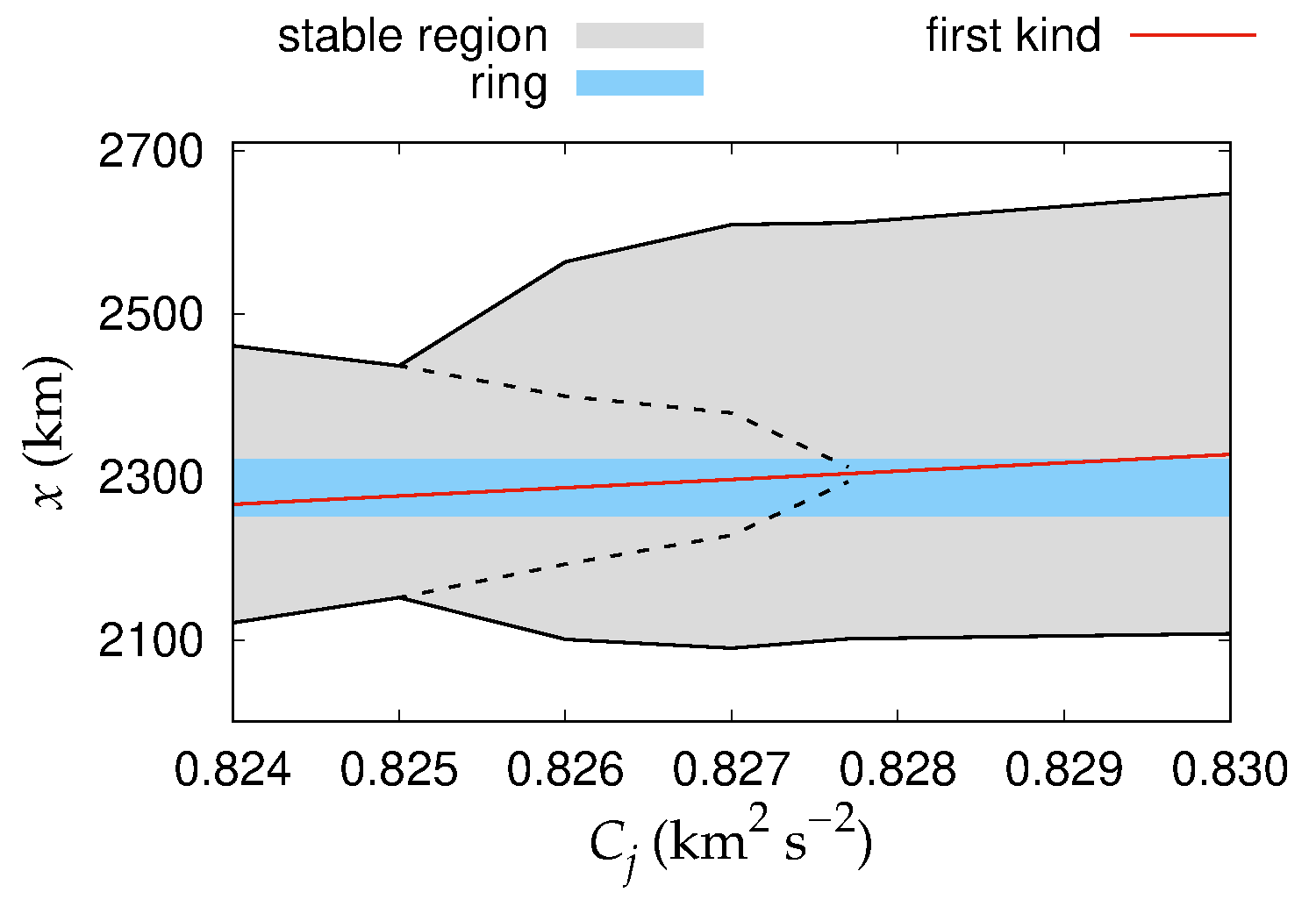}
\caption{Stable region associated with the periodic orbits of the first kind.  The stable region (gray) was determined from the Poincar\' e surfaces of 
section computing, for each Jacobi constant ($C_j$), the pair of $x$ values of the largest islands when $\dot x = 0$.  The dashed lines indicate the 
location of the separatrix of the 1:3 resonance, when it enters this stable region. The red line shows the values of ($C_j, x$) for the periodic orbits of 
the first kind. The blue region is just an indication of the location of the ring in terms of the $x$ range, not taking into account the values of $C_j$.}
\label{fig_axe}
\end{center}
\end{figure}

%%%%%%%%%%%%%%%%%%%%%%%%%%%%%%%%%%%%%%%%%%%%%%%%%%%%%%%%%%%%%%%%%%%%%%%%%%%%%%%%
\section{Final Comments}

In this paper, we presented a study on the dynamical structure in the region of the Haumea's ring.
Poincar\' e surfaces of section revealed that the 1:3 resonance is  doubled, presenting mirrored  pairs of periodic orbits.
The separatrix,  due to the resonance be doubled, generates a layer of chaotic region that  reduces the size of the stable regions produced by 
quasi-periodic orbits around the periodic resonant ones. 

An analysis of the semi-major axes and eccentricities of the 1:3 resonant periodic orbits showed that there is a minimum value of the eccentricity for 
 such orbits to exist. 
This minimum value showed to be too large for the radial location and size of the Haumea's ring.

Local viscous and/or self-gravity effects on the ring particles were not taken into account in the current study.
Having in mind that the Haumea's ring is only about 70 km wide, it should be extremely massive in order to reduce 
the particle's  eccentricity  produced by the 1:3 resonance. 

Collisions between the ring particles were also not considered in this work. 
They could allow  large orbital eccentricities of particles at the 1:3 resonance to be damped, and fit within the radial range of the ring.
Nevertheless, particles at the ring borders would  not remain confined, still needing a confining mechanism.
However, independently of collisions, particles associated with the first kind periodic orbits define regions of stability that fit very well in size and 
location that of the Haumea's ring. 
Therefore, this analysis suggests that the Haumea's ring is in a stable region associated with a first kind periodic orbit instead of the 1:3 resonance.

%%%%%%%%%%%%%%%%%%%%%%%%%%%%%%%%%%%%%%%%%%%%%%%%%%%%%%%%%%%%%%%%%%%%%%%%%%%%%%%%
\section{Acknowledgements}
This work was financed in part by Coordena\c c\~ ao de Aperfei\c coamento de Pessoal de N\'\i vel Superior - CAPES (Finance Code 001),  
CNPq (Procs. 312813/2013-9) and FAPESP (Procs. 2016/24561-0 and 2016/03727-7). 
These supports are gratefully acknowledged. 
The authors would like to thank Ernesto Vieira Neto for some criticism that  improved the work, and
Silvia Giuliatti Winter and Rafael Sfair for fruitful discussions.
The authors also thank the referee for his/her comments and questioning that helped to better understand some important issues of the paper.

%%%%%%%%%%%%%%%%%%%%%%%%%%%%%%%%%%%%%%%%%%%%%%%%%%%%%%%%%%%%%%%%%%%%%%%%%%%%%
\renewcommand{\refname}{REFERENCES}

\label{lastpage}

\end{document}